# Harnessing the Potential of Spatial Statistics for Spatial Omics Data with *pasta*

Martin Emons[1,†], Samuel Gunz[1,†], Helena L. Crowell[2], Izaskun Mallona[1], Malte Kuehl[3,4], Reinhard Furrer[5], and Mark D. Robinson[1,*]

[1]Department of Molecular Life Sciences and SIB Swiss Institute of Bioinformatics, University of Zurich, Zurich, Switzerland

[2]Centro Nacional de Análisis Genómico (CNAG), Barcelona, Spain

[3]Department of Clinical Medicine, Aarhus University, Aarhus, Denmark

[4]Department of Pathology, Aarhus University Hospital, Aarhus, Denmark

[5]Department of Mathematical Modeling and Machine Learning, University of Zurich, Zurich, Switzerland

[†]Equal contribution: both reserve the right to list themselves as first author; author order was determined by flipping a Swiss 5 franc coin.

[*]Correspondence to: mark.robinson@mls.uzh.ch

**Abstract**

Spatial omics allow for the molecular characterization of cells in their spatial context. Notably, the two main technological streams, imaging-based and high-throughput sequencing-based, give rise to very different data modalities. The characteristics of the two data types are well known in spatial statistics as *point patterns* and *lattice data*. In this perspective, we show the versatility of spatial statistics to quantify biological phenomena from local gene expression to tissue organization. As an example, we describe how to use exploratory metrics to address scientific questions in breast cancer, including cellular co-localization and gene co-expression analysis. We discuss technical concepts like window sampling, homogeneity and weight matrix construction and show their importance. We also provide *pasta* (https://robinsonlabuzh.github.io/pasta), an extensive analysis vignette for spatial statistics both using R and Python packages with further biology-driven applications.

# Introduction

## Molecular profiling at spatial resolution

Molecular profiling of cells in bulk or at the single-cell level can be accomplished by dissociating organs or tissues. Dissociation may select against certain cell types [28, 29] and results in the loss of the spatial organization. This limitation is attenuated with spatial profiling techniques, which preserve the cellular spatial organization and range from spatial proteomics [36, 45, 41, 17, 83], to spatial transcriptomics based on fluorescent *in situ* hybridisation [47, 23, 44] or on sequencing [79, 60, 75, 80, 22], to spatial epigenomics [50] and spatial metabolomics [3, 94], or multi-omics [85]. The technologies and their application in biological research have been the topic of various detailed reviews [10, 55, 70, 58, 62, 18].

Spatial data modalities have been present in other fields for decades. They originated in mining and forestry and have expanded to other fields such as geography or ecology, and have been approached by spatial statistics methods [96, 25, 11, 66]. In this work, we focus on the concepts and application of exploratory statistical approaches for molecular spatial data; our overview also centres on key differences between data modalities that result from current spatial profiling technologies and the different types of questions that can be answered with spatial statistics. In this regard, we complement the ongoing discussion on the challenges in applying spatial statistics to spatial (transcript)omics data [95, 59, 37].

## Data modality is relevant to spatial data analysis

Most spatial omics assays can be classified as either high-throughput sequencing (HTS)-based or imaging-based. In HTS-based approaches, positional information is recorded by barcoding a predetermined array of spots or beads. Imaging-based approaches, however, either target the molecules of interest with fluorescent probes, ablate regions stained with a cocktail of antibodies via metal tag readouts, or amplify and sequence target sequences *in situ*. Several technologies are emerging, but the main trade-offs stem from the resolution, number of features, and sensitivity of the readout [70, 64]. For example, HTS-based approaches aim to profile the entire transcriptome (i.e., untargeted), but come at a resolution determined by the spot size. On the other hand, imaging-based approaches typically describe a lower number of features (i.e., targeted) but show a higher resolution. Note also that the landscape of spatially-resolved molecular technologies is rapidly changing, mostly in the direction of higher resolution for HTS-based methods and higher-plex marker panels for imaging-based approaches (Figure 1A; Supplementary Table S1) [55, 70, 69].

In terms of data analysis, HTS- and imaging-based outputs are quite distinct. HTS-based approaches collect data along regularly spaced spots or beads resulting in a so-called *regular lattice*. In contrast, imaging-based approaches measure features at exact locations that can be assumed to have originated from a random (stochastic)process known as a *point process* [70], leading to so-called point patterns. Although a random process might not be the correct assumption in biological development, the representation of point patterns can still be useful when correctly interpreted as we will show later. Thus, we will distinguish between lattice-based and point pattern-based spatial omics data. While the classification is often on the technology (c.f., imaging-based versus HTS-based), we argue that the distinction should lie at the data representation level [70]. This is important, since there are technologies that can be represented as both point patterns and lattice data depending on the resolution and data processing. For example, imaging-based data can be processed to describe the location of single transcripts at subcellular resolution (*point pattern*) while also identifying in-dividual cells as either an *irregular lattice* when using the cell segmentation as the lattice structure or as a *point pattern* when considering cell centroids. In general, if events of a point pattern are aggregated into specified regions (e.g., transcripts per cell from segmentation), we end up with lattice data (Figure 1B) [25, 66, 24].

## Lattice and point pattern analysis represent two overlapping streams of spatial statistics

Since we can represent some spatial omics datasets as either lattice or point pattern, it is important to further understand the assumptions underlying common data analysis strategies. On the one hand, point pattern analysis assumes that the point locations were generated by a random process, where each point is an event itself, a so-called *event-based* view of the data; its goal is to study the data generating process [11].For example, if we are interested in the distribution of one cell type in an image, we would use point pattern methods to quantify this arrangement. On the other hand, one could assume that the locations are fixed and known at the time of sampling and study the associated features at each location via an *observation-based* view of the data, while recognizing that the observations in the lattice are not independent measurements due to their spatial dependence [8]. For example, in spot-based technologies, locations are predefined. If we want to study a gene expression pattern in such an assay, we should account for the spatial dependence between spots (Figure 1C).

In addition, characterizing the *marks* (variable) of interest is also important. We consider two broad types of marks associated with spatial locations: *categorical* (qualitative, e.g. cell types or spatial domains) and *numerical* (quantitative, e.g. gene expression). In terms of dimensionality, marks can be analyzed one- (*univariate*), two- (*bivariate*) or three or more at a time (*multivariate* analysis). Figure 1C summarizes options to analyze each combination of the type of mark and its dimensionality. Defining characteristics of marks is relevant to choose appropriate metrics for the biological question in mind; e.g., infiltration of cell type A into a region dominated by cell type B is a question involving a categorical bivariate measure (two cell types involved) and can be analyzed as a type of co-localization analysis between type A and B cells.

Notably, there are more methods for the analysis of categorical marks with point patterns. For numerical marks, more options exist in lattice data analysis (Figure 1). This should serve as an indication of which stream can be useful, if the data modality from the experimental assay allows for both point pattern and lattice data analysis.

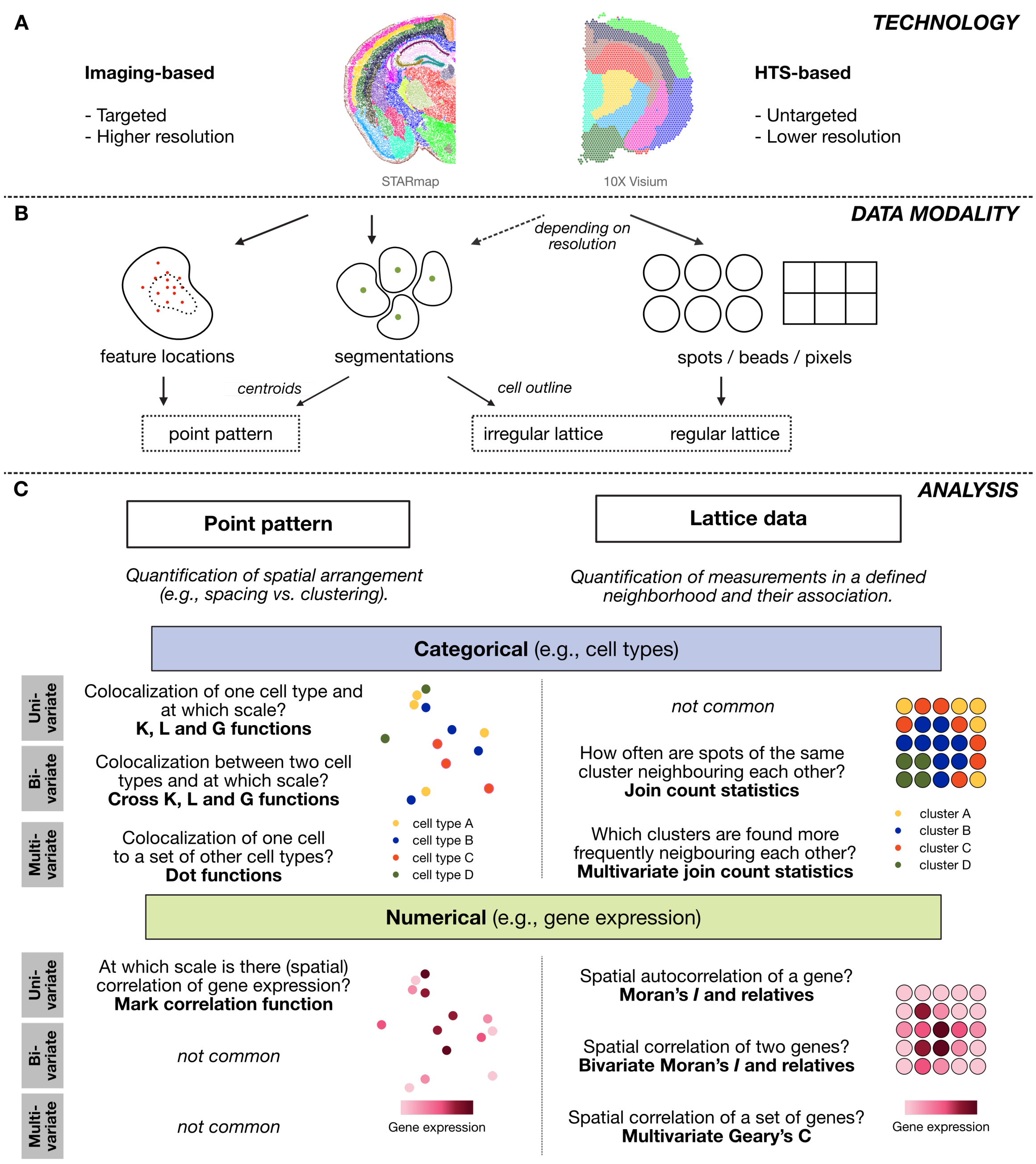

*(Figure legend on next page.)*

Here, we will explore the application of spatial statistics to spatially-resolved omics data guided by concrete biological questions (Figure 1). We will further focus on the two main streams of data analysis, point pattern-based and lattice-based methods, describing their strengths and limitations. Finally, we will give an outlook on the challenges and research gaps in the field of spatial data analysis for omics data.

Figure 1: Overview of the two streams of spatial omics data and corresponding analysis strategies. **A**) *Technology*: Spatial omics technologies can be divided into two major streams: imaging-based and high-throughput sequencing (HTS)-based. Examples of both imaging-based (STARmap PLUS [77]; left) and HTS-based (10x Visium [1]; right) datasets are shown. **B**) *Data modality*: these technology streams lead to distinct data modalities. Imaging-based omics can represent both randomly-generated point patterns as well as irregular lattices. Most HTS-based data, on the other hand, can be interpreted as a regular lattice due to the regularity of the sampling locations; approaches are emerging that could allow high-resolution HTS-based data to be segmented into cells [68]. **C**) *Analysis*: depending on the data representation and the feature of interest, different approaches of data analysis are available to the analyst. Depicted are exemplary questions divided by feature of interest and type of analysis and corresponding methods for analysis.

# Biological applications of spatial statistics for spatial omics

## Spatial autocorrelation reveals triple positive regions in breast tumours

We re-analyzed an imaging-based human breast cancer dataset [44], where the cell-level data can be considered as both a point pattern or lattice data (Figure 1). Figure 2A shows the H&E image of the breast cancer section.

As the first feature of interest, we consider the measured gene expression, which is a numerical, continuous (cell-level) "mark". As we are interested in relationships within the local neighbourhood of the cells rather than effects at specific locations or distances, lattice data is the appropriate analysis method (Figure 1).

In their publication, Janesick *et al.* [44] investigate the expression of three clinically relevant hormone receptor genes: Estrogen receptor (*ESR1*), progesterone receptor (*PGR*) and human epidermal growth factor receptor 2 (*ERBB2*). The presence or absence of these receptors (indicated with +/-) can inform treatment strategies [44]. The expression pattern of each receptor is quite distinct, with *ERBB2* being more highly expressed compared to both *ESR1* and *PGR* (Figure S1A-C). Lattice data analysis enables us to use a locally-scaled measure of expression to distinguish high and low expression regions. Since we are interested in localized hormone receptor-positive regions, we define the neighbourhood as the six nearest neighbours around each cell.

Univariate spatial autocorrelation allows us to identify regions where the expression of the hormone receptor genes is correlated among cells in their neighbourhood. Since autocorrelation measures such as local Moran's $I$ [57, 6] are non-directional (e.g., local autocorrelation can be both high in regions where expression among neighbours is high or low as the focus in on similarity; for more details, see section "Spatial autocorrelation is a measure of similarity"), we rely on Moran's scatter plot to identify the type of interaction [7]. In regions with "high-high" interaction, the expression among neighbours is both similar and high, which can be interpreted as a region positive for one of the receptors (Supplementary Figure S1D-F). Figure 2B shows the union of the individual positive univariate classifications of the three receptor genes and indicates single, double and triple positive receptor regions. One DCIS region (indicated by arrow) is lined with predominantly triple positive cells, but some local heterogeneity with double and single positive cells is apparent.

We also applied bi- or multivariate spatial correlation measures to classify the different regions. Similar to spatial autocorrelation, bivariate autocorrelation measures the similarity of expression of two genes in a local neighbourhood. Supplementary Figure S2A-C show bivariate local Lee's $L$ [48] correlation measures for the pairs of receptors *ERBB2-ESR1*, *ERBB2-PGR* and *ESR1-PGR*, respectively; regions with high values indicate regions where expression of both receptor genes is similar ("high-high" or "low-low"). The different scales of the local Lee's L $L$ values are the result of the different expression scales (c.f., Supplementary Figure S1A-C). A comparison of gene expression values and the local Lee's $L$ value allows us to identify double positive regions.

Multivariate spatial correlation metrics extend the concept of spatial correlation to more than two genes. The main difficulty is finding an appropriate metric to measure similarity of more than two marks. In multivariate Geary's $c$, the respective metric is the weighted average of the squared differences of the expression vectors of the three genes. Supplementary Figure S2D-F shows an analysis of multivariate local Geary's $c$ [4] for the three hormone receptors. Cells with a high local Geary's $c$ value exhibit a gene expression profile that is highly dissimilar (squared difference is high) to their local neighbours and indicates local heterogeneity. Interestingly, the triple positive region identified using Moran's scatter plot shows high multivariate Geary's $c$ values. At first, this seems counterfactual to what we have identified using the Moran's scatter plot. A closer look reveals that the region is very heterogeneous with a lot of single and double-positive regions with varying scales of expression of the individual genes. Generally, there is a difficulty in interpreting multivariate autocorrelation measures as they are not simply the sum of the respective univariate counterparts and should therefore be mainly used to identify interesting regions (c.f., Anselin *et al.* [4]).

In summary, we have investigated the expression patterns of the three clinically-relevant receptor genes, quantifying single-, double- and triple-positive receptor regions in the tumour using lattice data analysis. Spatial statistics was able to recapitulate the original findings in Janesick *et al.* [44] (Figure 5), and in addition, spatial exploratory statistics provide a quantification of the strength of the various spatial correlations as well as an indication of the significance of the effects.

## Point pattern analysis quantifies different degrees of tumour invasion into ductal carcinomas

Another analysis to consider is the exploration of the spatial distribution of cell types in the Janesick *et al.* [44] dataset. Since cell types are categorical features, point pattern analysis is a powerful choice [11] (Figure 1). Here we consider the DCIS 1, DCIS 2 and invasive tumour cell types and aggregate all other cell types into the "Other" category (Figure 3A). In this example, we calculate Besag's cross $L$ function to quantify the co-localization of the three cell types in space [14, 73]. Note that the observation window was restricted to the region where most of the cells were found to account for the underlying inhomogeneous distribution of cells (c.f., Figure 3). There are other options to account for this inhomogeneity, which are compared in Figure S3. In general, Besag's $L$ calculates the average number of neighbours of a certain cell type as a function of the radius. This can used to estimate the magnitude of co-occurrence relative to a scenario where all cells are randomly distributed in the observation window. In Besag's $L$, this corresponds to a value of 0 across all radii. Positive deviation indicates more cells are observed than expected by chance, implying clustering; negative deviation indicates less cells observed than expected by chance, implying spacing of the cells.

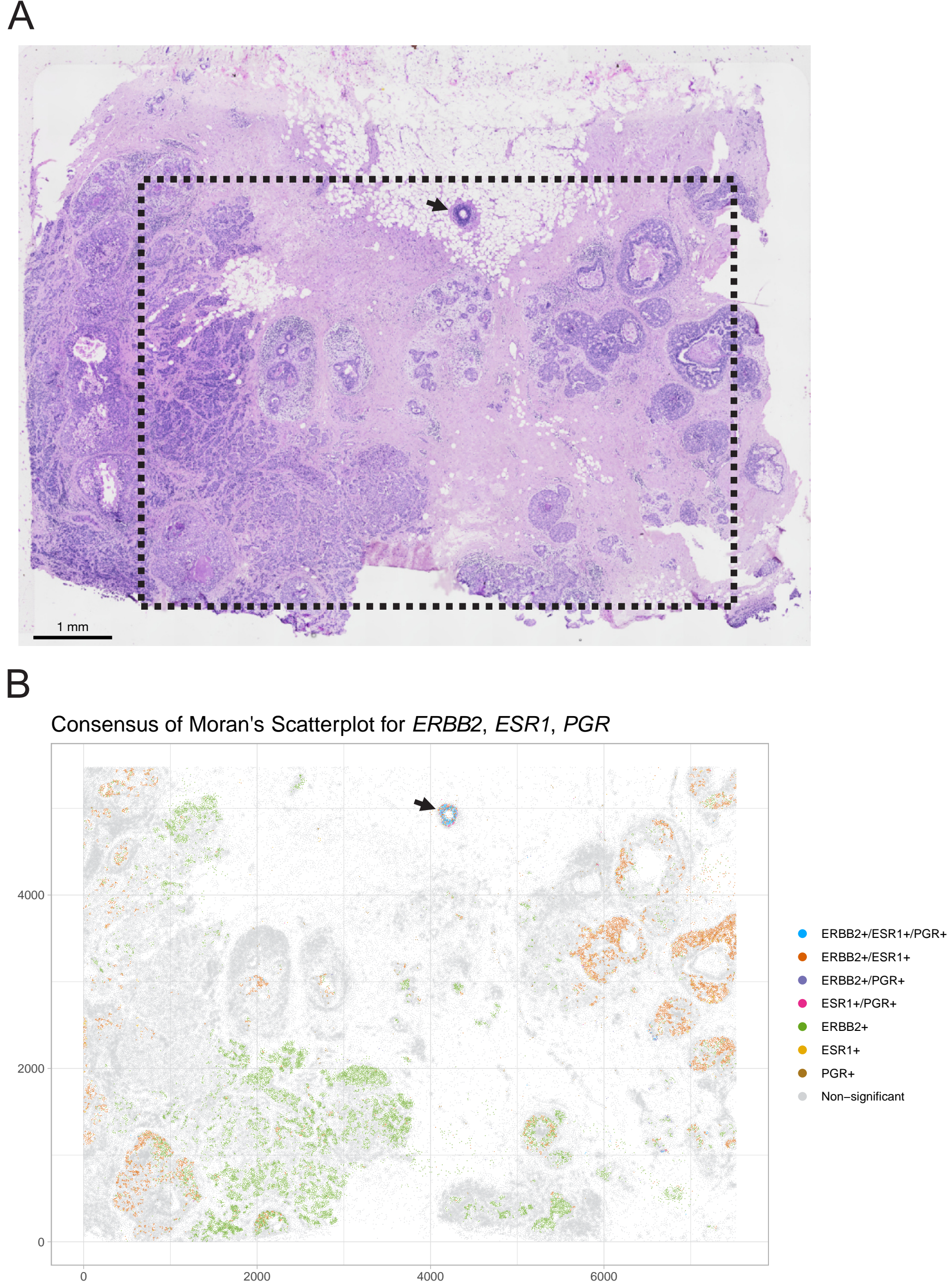

Figure 2: Analysis of three receptor genes (*ERBB2*, *ESR1* and *PGR*) in a Xenium human breast cancer data set [44]. **A**) H&E image of a breast cancer section. The black dashed line indicates the region profiled by Xenium (hand annotated). **B**) Moran's scatter plot [7] highlighting the status (+/-) of each cell for the receptors *ERBB2*, *ESR1* and *PGR*. Double and triple positive results were derived from the union of the univariate "high-high" clusters using Moran's scatter plot and significance scores from local Moran's $I$. The triple positive region is indicated with an arrow in A) and B).

First, we consider univariate co-localization, i.e., whether cells are found more often than expected in the vicinity of other cells from the same type. Figure 3B shows Besag's $L$ curve plots for each pair of cell types. We note that all cell types exhibit curves above the (dashed black) reference line, which means that we observe, on average, more cells of the same type within a certain radius than expected by chance. This represents a higher degree of clustering than a fully random, or completely spatial random (CSR), point pattern (see Technical description below; c.f., box "Point Processes"). Thus, all three cell types have a propensity higher than expected by chance to occur in the neighbourhood of their own type (Figure 3A) and this clustering is strong across all scales (or radii, $r$).

Besag's $L$ can quantify bivariate interactions, i.e., the co-localization of a pair of cell types, as shown in Figure 3B. We note that the combination of DCIS 1 and DCIS 2 is below the dashed CSR line. This means that we observe less cells of the other type within a specific radius than expected by chance. This can be interpreted such that there is, on average, *spacing* between the respective cell types. Of interest is the co-localization of the two DCIS subtypes and invasive tumour cells: DCIS 1 cells and invasive tumour cells are strongly spaced, thus clearly co-occur less frequently than expected at random. On the other hand, DCIS 2 cells and invasive tumour cells are close to the CRS line which means they occur as often as expected by chance, an indication of mixing of the two cells types in space.

In summary, we have recapitulated the main findings on DCIS cells presented in Janesick *et al.*, Figure 4, using point pattern analysis [44]. DCIS subtypes show a different degree of co-localization with invasive tumour cells, with DCIS 1 regions showing no co-localization with invasive tumour while DCIS 2 exhibits a degree of co-localization with invasive tumour cells that can be expected by chance. Using spatial point pattern analysis, we have systematically quantified the co-localization of three cell types in breast cancer.

## Further analyses using spatial statistics

In addition to the analyses described above, spatial statistics could be used for various other analyses.

Univariate global spatial autocorrelation methods such as Moran's $I$ can be used to identify the strength of spatial correlation across all genes in the panel. This information can be useful to compare spatial variability of different genes within one sample and the specified neighbourhood. In addition, by constructing a spatial weight matrix based on contiguous neighbours (i.e., cells in physical contact with each other), we can identify pairs of ligand-receptor genes via quantification of their bivariate spatial correlation [49]. Alternatively, adjusting the weight matrix to neighbours based on a certain distance, we can investigate paracrine signalling up to the distance specified.

Investigating the spatial arrangement of cells with point pattern analysis, we could further use mark correlation functions to quantify the scale of spatial correlation of the expression of a gene of interest. In addition, we could profile the co-localization of cell types using nearest-neighbour approaches such as the nearest-neighbour distance and orientation. These metrics would provide us with a more quantitative readout of the spatial organization than we can identify by eye. "Dot" (or "$i$-to-any") functions would allow us to make statements about how one cell type is localized with respect to a set of other cell types.

Moreover, we could also focus our analyses to regions of interest such as the single-, double- and triple-positive tumour regions to make statements about localized gene expression and the arrangement of cell types within these regions. The observation scale is very important in spatial analysis as it will impact the interpretation of spatial relationship, as we will further explain below.

# Technical considerations

## Inference about the point process

In the analysis of point patterns, the goal is to make inferences on the point process that generated the data, not on the patterns themselves [11, p. 127]. When considering a point pattern with many cell types, there are two possible formulations. In the first variant, we assume that cells in a section depend on each other so their distribution is due to one overarching biological process; this is referred to as a *multitype* view. In contrast, when we consider the $m$ patterns are created by $m$ point processes, we assume that these processes can be individually analyzed; this represents the *multivariable* viewpoint (change in terminology in comparison to [11]). Generally, it is not easy to define whether the entire pattern of several cell types should be represented as a single point process (multitype analysis). Notably, analyzing cell types individually and considering them as separate point processes might underestimate dependencies among the cells (multivariable analysis) [11, pp. 565 ff.].

## Point pattern analysis quantifies spatial interactions across scales

A central idea in point pattern analysis is that of an $r$-neighbourhood. The $r$-neighbourhood describes the spatial dependence in a disk with radius $r$. In such a neighbourhood, different aspects can be quantified, leading to different dependent variables per function. The expected number without any spatial interaction (CSR or Poisson process) can be calculated in the $r$-neighbourhood. Positive deviations from the CSR value indicate clustering whereas negative deviations indicate spacing [11, p. 204]. Therefore, point pattern metrics are functions that quantify spatial dependence across scales.

The two most well-known classes in point pattern analysis are *correlation* and *spacing* functions. The two approaches are highly complementary and there is a duality between them [11, p. 255]. Correlation is usually assessed with Ripley's $K$ function or its variance-stabilized version, Besag's

A

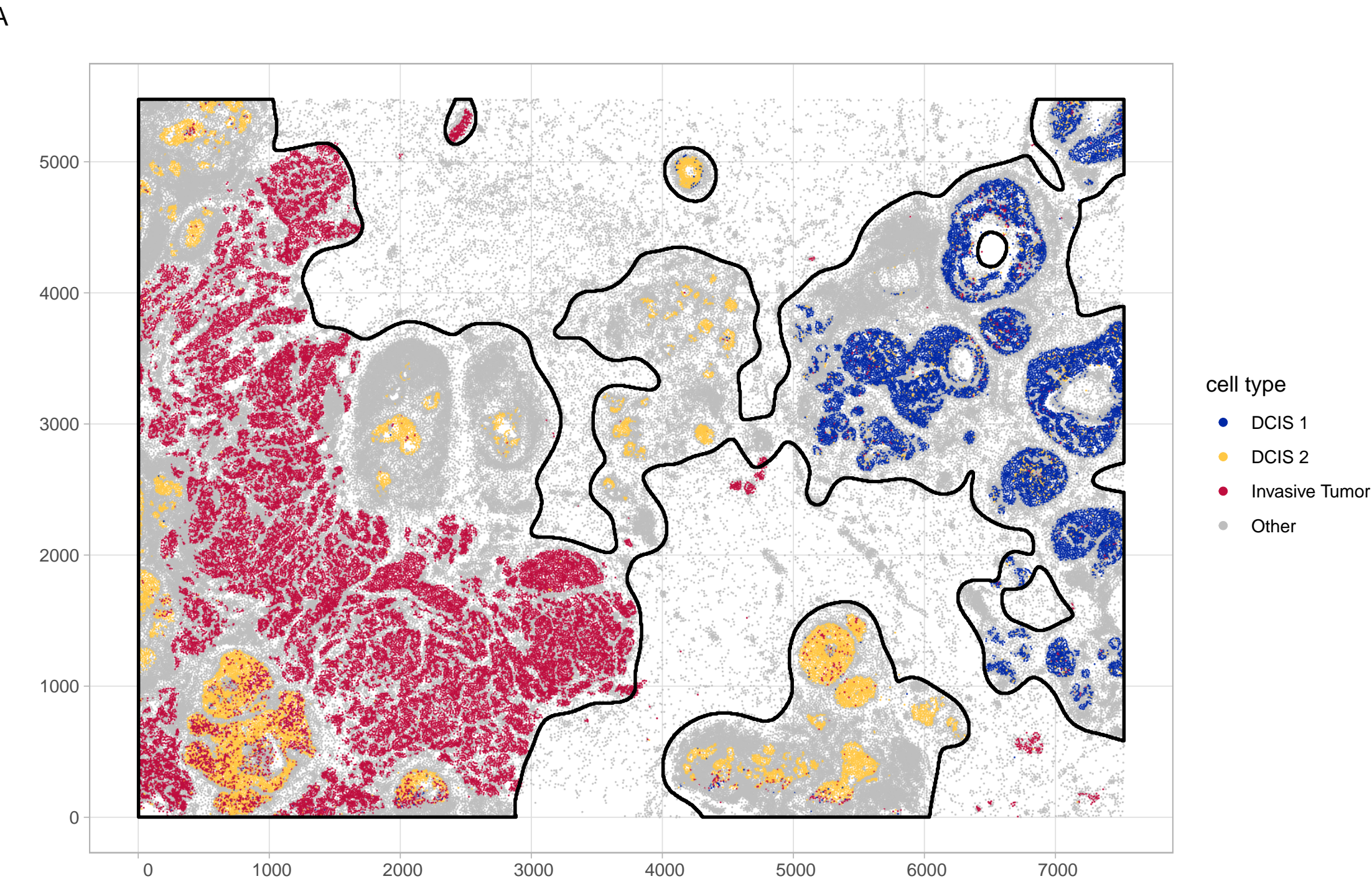

B

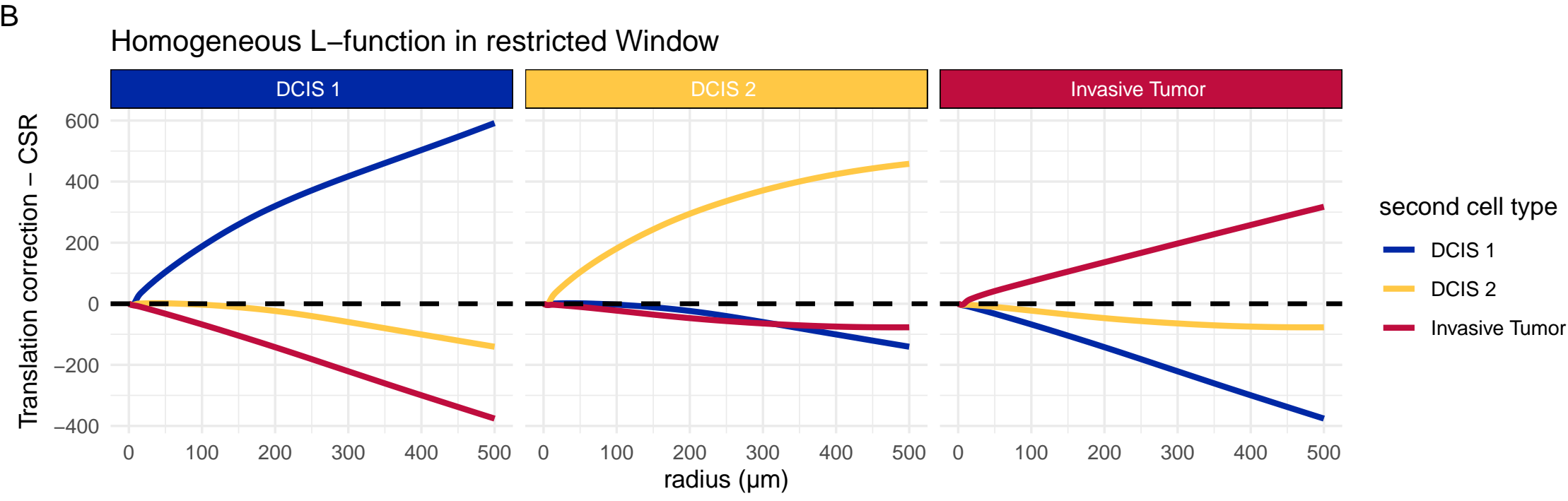

*(Figure legend on next page.)*

Figure 3: Point pattern analysis of different cell types in a Xenium human breast cancer data set [44]. **A**) A Xenium human breast cancer data set. The labels were obtained from 10x Genomics by supervised clustering of the Xenium data. The plot is coloured by four categories: Three cell types (DCIS 1, DICS 2 and invasive tumour) and a bin for all remaining cell types as "Other". The coordinates are indicated in micrometers [44]. The observation window was determined using a threshold on the unmarked point pattern intensity and is indicated in black [40]. **B**) Besag's $L$ function for all pairwise combination of the three cell types: DCIS 1, DICS 2 and invasive tumour computed in the restricted observation window indicated in A). The $x$ axis represents the increasing radius $r$ of the expanding circle, the $y$ axis represents a measure of how many points fall in this radius, corrected for edge-effects. The black dashed line indicates Besag's $L$ of a CSR process. Positive deviation from the CSR line indicates clustering, negative deviation indicates spacing.

$L$. Here, the number of points falling in an $r$-neighbourhood is quantified. These raw counts are corrected for varying intensities as well as edge effects (see Box Point Processes) [74, 14][11, p. 203 ff.]. Spacing can be assessed with the nearest-neighbour function $G$. Here, the probability of finding the nearest-neighbour of a point in an $r$-neighbourhood is calculated [11, p. 261 ff.]. Both spacing and correlation quantify different aspects of a point pattern and a spatial analysis should consider both complementary options [11, p. 295 ].

## Window sampling and small world impact representative sample collection

Many imaging platforms measure patterns in a field of view (FOV) and collect several such FOVs per sample. In some technologies, there are no gaps between the individual FOVs so they can be assembled into one image. If there are gaps, consecutive FOVs can either be stitched together (which can be computationally complex) or analyzed individually [10, 56]. When analyzing individual FOVs, not the entire pattern but rather a subset of the full pattern is observed: this is so-called *window sampling* because the window is a sample of the full point pattern. A related concept is the *small world model*, which describes that points can only be observed in a finite world and not beyond these boundaries [11, pp. 143ff., p. 560]. Often, the distinction between the window sampling and small world model concepts is not clear in histological sections. An example is the arrangement of the epithelium, which can can be imaged with several FOVs (window sampling) but there are no cells expected outside of the epithelial boundary (small world) [32]. Therefore, in spatial omics, we often encounter a mixture of these two scenarios.Obtaining a representative sample of a point pattern is also a challenge. In spatial omics, we are often provided with FOVs that an experimentalist has selected from a much larger region or that are the result of the data acquisition (e.g., FOVs in CosMx [43]). Such a sample might have different characteristics compared to the point pattern of the larger region, as FOVs are sometimes selected based on morphological properties (e.g., H&E staining). In cases where we have a small world scenario, we would want to limit our observation window to the full region where the points can occur (c.f., Supplementary Figure S3).

## Importance of scale and uniformity in point pattern analysis

Apart from the window of measurement, we need to make assumptions on the statistical properties of the point pattern, including, most importantly, whether points can be considered *homogeneous* or not. Homogeneity (see Box Point Processes) assumes that the number of points in a given region is proportional to its area, i.e., homogeneity refers to a uniform *intensity* of points across the window of measurement. Otherwise, the point process is inhomogeneous [11, p.132-133]. This difference has important implications for the analysis of a point process. For example, when adjusting for varying local intensities, a process could be interpreted as inhomogeneous rather than genuinely clustered. This is called the confounding between intensity and interaction (see Box Point Processes) [11, pp.151-152]. Figure 4A shows a single FOV with islets from the human pancreas acquired using IMC [27]. An analysis considering the entire FOV will tell us about the processes that gave rise to the spatial distribution of the endocrine pancreas (islets). As we see in Figure 4B, a Ripley's $K$-function under a homogeneity assumption indicates that in this observation window, endocrine cells are clustered within pancreatic islets (the line lies above the theoretical CSR line). Figure 4B corrects for inhomogeneity of islets cells in the total fov and indicates CSR at $r < 75$ µm. Note that the calculation of the inhomogeneous version varies with the estimation of the intensity function and the border correction (c.f., Supplementary Figure S5), especially when the data is not correlation-stationary [11, pp. 321ff.]. However, there might also be a relevant biological scale where we subset our observation window to the islets themselves (Figure 4D). In this case, Ripley's $K$-function indicates a homogeneous distribution of the islet cells within the observation window (Figure 4E, the line follows the theoretical CSR line). Overall, the correct scale needs to be chosen in accordance with the research question, since the interpretation depends on it.

**Point Processes**

**Intensity** The locally expected number of points in a given area is called intensity. It represents a first moment property, since it is related to the expected number of points. The mean intensity of a homogeneous process can be defined as $\bar{\lambda} = \frac{n}{|W|}$ [11, p. 149, p. 160].

**Homogeneity** If the expected number of points falling into an arbitrary region $B$ is proportional to the area $|B|$ (with proportionality constant $\lambda$), the process is called homogeneous [11, p. 132].

**Independence** Independence of region counts implies that the number of points in disjoint regions $A$ and $B$ are independent random variables [11, pp. 132-135].

**Complete spatial randomness** If a process is both homogeneous and independent (see above), the process is said to be completely spatially random (CSR) [11, pp. 132].

**Edge effects** A consequence of window sampling. Since the window $W$ does only contain a subset of the point process, the statistics could be biased along the edges of $W$ (e.g., nearest neighbour outside of $W$) [11, pp. 213 ff.].

**Confounding between intensity and interaction** Often, it is hard to distinguish inhomogeneous intensity from genuine interaction of points (clustering). Patterns can arise from an inhomogeneous process or a clustered process and it is not straightforward to distinguish them: one is governed by different (inhomogeneous) intensities whereas the other is due to interactions of points [11, pp. 151 ff.]. One example of this confounding is cells of different areas being treated as points instead of considering their entire volume [11, p. 210].

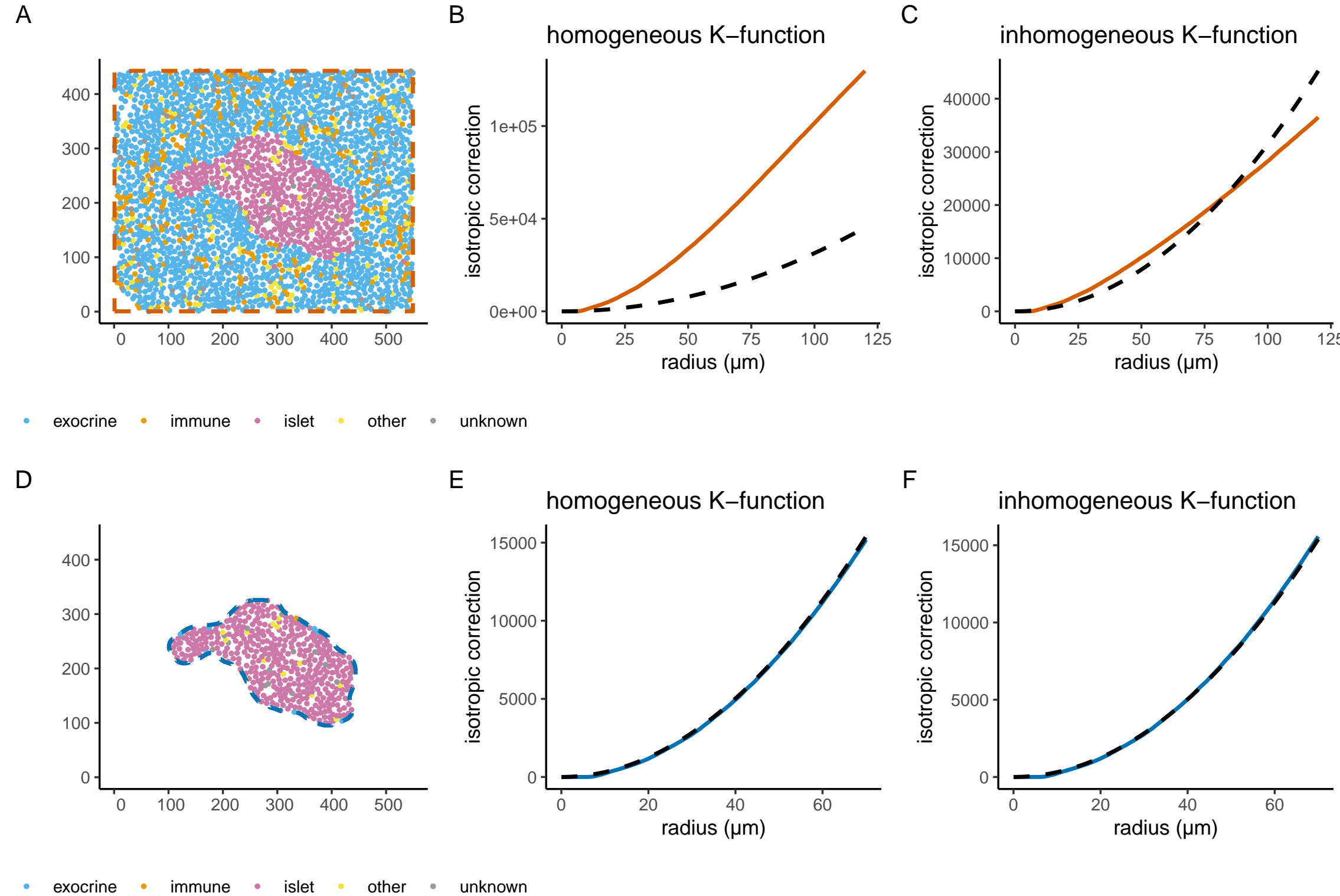

Figure 4: **A**) Single image of an IMC proteomics dataset showing islets in the human pancreas [27]. The points are the centroids of cells coloured by cell type category. The coloured dashed line indicates that the analysis window is set to be the entire FOV. Axes on the µm scale. **B**) Global analysis of islet cells using a homogeneous $K$-function indicates strong clustering of islet cells **C**) Global analysis of islet cells using an inhomogeneous $K$-function indicates behaviour close to CSR of islet cells at $r < 75$ µm when accounting for inhomogeneity. **D**) The same image as in A) but the analysis window (blue dashed line) is constrained to the islet. The observation window was determined using a threshold on the density of islet cells [40]. **E**) Local analysis of islet cells using a homogeneous $K$-function show no clustering of islet cells in the window **F**) local analysis using an inhomogeneous $K$-function reveal as well no clustering of islet cells. Coordinates in A/D) correspond to $r$ values in the plots. The black dashed line in the plots corresponds to a CSR distribution (cells randomly distributed).

## Regular and irregular lattices

Several HTS-based spatial transcriptomics technologies produce data on a regular lattice that gives evenly-spaced spots or beads of uniform size and shape, whereas imaging-based technologies yield irregular lattice structures with variable cell sizes and shapes, and non-uniform spacing [96, pp.

321ff.] [5]. Once we have recorded the outline and arrangement of the spatial units in the lattice, we can specify the strength of spatial relationships between each unit. Each pair of units is assigned a weight: the stronger the connection between units, the higher the weight. The collection of all weights between locations forms the weight matrix (see Box Lattice Data).

## Spatial autocorrelation is a measure of similarity

A common exploratory analysis for lattice data is based on the concept of spatial autocorrelation. Spatial autocorrelation measures the degree of association of features, e.g., the expression of a gene, that are assumed to be dependent in space. For each combination of spatial locations, a measure of association is calculated and scaled by the weight of the connection [5]. This approach considers both the proximity (via the weight matrix) and the characteristics of the locations (via the metric) [96, pp. 327ff.] [66, pp. 209ff.] [35]. When interpreting (local) autocorrelation measures, it is important to consider both the effect size estimates and the significance level. Since the significance level is calculated for each spot separately, it is recommended to adjust for multiple testing (e.g., Benjamini and Hochberg [13]). Overall, local Moran's $I$ statistics reveal locations that have similar values to their neighbours. However, in mouse brain, regions with the highest and most significant local Moran's $I$ value lie where expression of gene *Nrgn* is very low among neighbours (Figure 5A-B). Notably, the local Moran's $I$ measure is both dependent on the value of the log-transformed counts and the similarity among neighbours (c.f., Supplementary Figure S4A-B).Metrics such as the Moran's scatter plot can be useful to classify the type of interaction [7]. The Moran's scatter plot compares the values of each spot with the average of it's neighbours as defined in the weight matrix. This information can be used to classify each point into respective categories (c.f., Supplementary Figure S4E)). For example, in Figure 5 C the clusters denoted "high-high" refer to regions where high expression values are surrounded by high values, while "low-low" clusters indicate that low values are surrounded by low values (for which the local Moran's $I$ value can be high as well). "High-low" and "low-high" values denote heterogeneous regions.

## Definition of the neighbourhood is critical in lattice data analysis

The construction of the weight matrix is critical for all downstream analyses as it encodes the spatial relationship between the units in the lattice [96, pp. 321ff.]. Several construction strategies exist: contiguity-based (i.e., in direct contact), graph-based, distance-based, or higher-order neighbours [34, 96, 66]. Neighbours based on graph adjacency or on distance [66, pp. 191 ff.] are also used; the open question is which is best suited for spatial omics data. If one uses contiguity-based neighbours in imaging-based spatial transcriptomics, results ultimately depend heavily on the accuracy of the cell segmentation, which is known to be challenging [81, 39, 67, 9, 87]. If neighbours beyond the adjacent ones are used, the scale (e.g., at which cells interact) needs to be specified. Figure 5C-D show the difference between the local Moran's $I$ calculation when based on contiguity-based neighbours in D, the 10 nearest neighbours in E or neighbours within a 1000 pixel distance ($\sim 180\,\mu m$) in F. While the overall differences are small, some cells show different local Moran's $I$ values. Smoothing of the local Moran's $I$ values occurs when more neighbours are considered (Supplementary Figure S4C). Furthermore, Supplementary Figure S4A and B show that for some cells, no contiguous neighbours were found, which results in a zero estimate of the local Moran's $I$. Cells do not function in isolation but form complex anatomical structures including extracellular components. Whether analyses improve when weight matrix construction takes such anatomical structures or regions into account is to be determined. Furthermore, it remains to be investigated how much the construction of the weight matrix influences downstream analyses in spatial omics data.

### Lattice Data

**Lattice** A lattice or grid is composed of individual spatial units $D = \{A_1, A_2, ..., A_n\}$ where these units do not overlap. The data is then a realization of a random variable along the lattice $Y_i = Y(A_i)$ [96].

**Regular and irregular lattices** In a regular lattice, all spatial units have the same size, shape, and the observations are placed on a regular grid. If this is not the case, the lattice is irregular. Observations that follow the outline of natural objects are usually irregular lattices (e.g., cells in a tissue) [96].

**Weight matrix** The weight matrix $W = w_{ij}$ defines the spatial relationships between units $A_i$ and $A_j$. The neighbourhood matrix is a special case of a weight matrix, where all entries are 1 (direct neighbour, connection) or 0 (not a neighbour, no connection) [96].

**Spatial Autocorrelation** Spatial autocorrelation measures take the form $a_{ij}U_{ij}$, which uses a similarity measure $U_{ij}$ weighted by the strength of the connection $a_{ij}$ [35, 66].

**Global vs. Local spatial autocorrelation** Global spatial autocorrelation measures estimate the average level of spatial autocorrelation across the observations at all locations $\sum_i \sum_j w_{ij}U_{ij}$. Local measures or Local Indicators of Spatial Association (LISA) give us information about the statistic at each location. The global statistic can be seen as weighted sum of its respective local statistics. The concept of measuring local associations exists both in point pattern (c.f., Figure 4E) and in lattice data analysis (c.f., Figure 5B) [6].

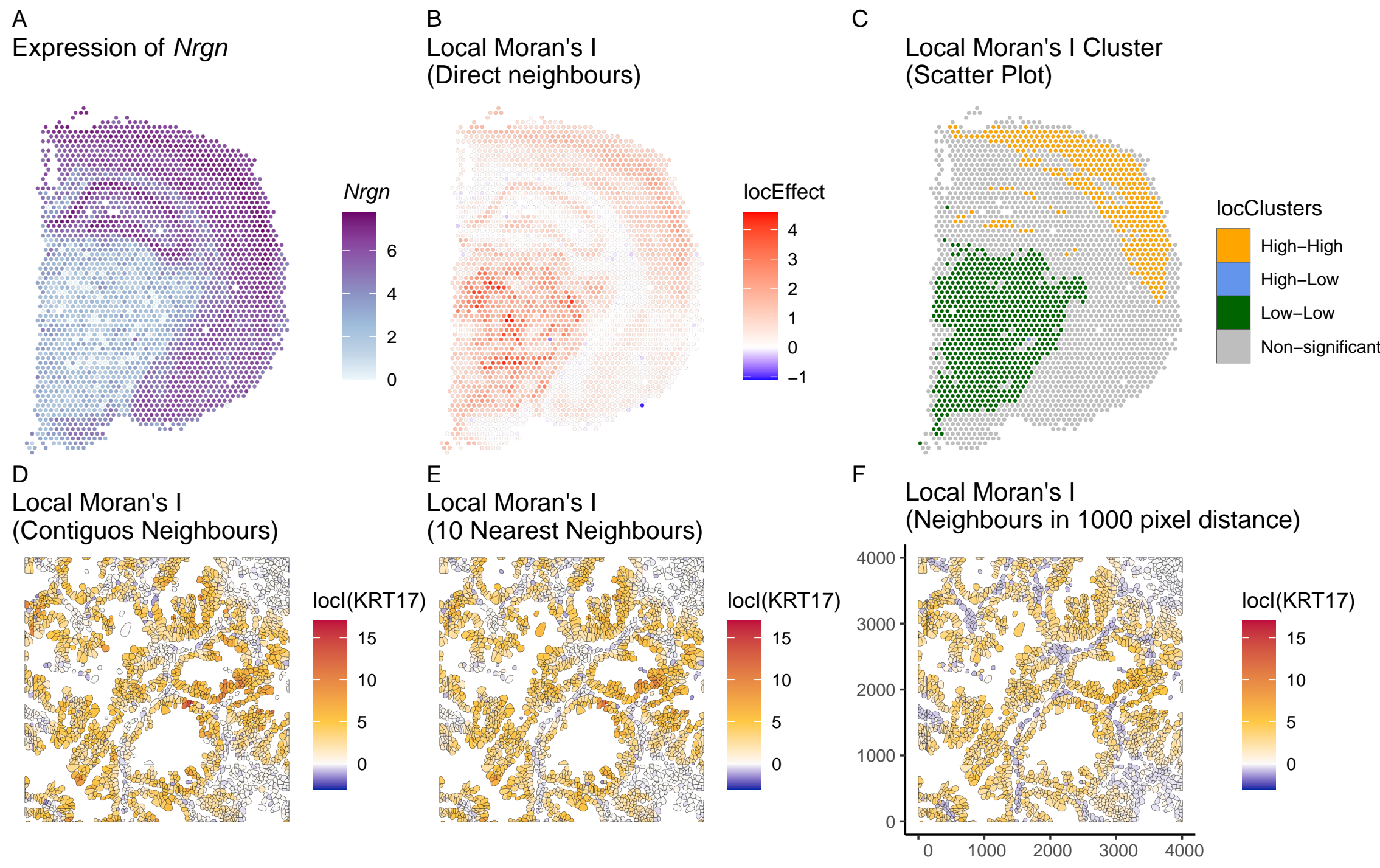

Figure 5: Spatial autocorrelation analysis. Panel **A**) shows log-transformed counts of *Nrgn* expression in the Visium mouse coronal brain section data [1]. We note regions with no expression (white), regions with low expression (blue) and regions with high expression (violet). **B**) shows the local Moran's $I$ values calculated based on the values in A). Local Moran's $I$ is high in regions with low and high expression of *Nrgn*. **C**) shows the clusters of Moran's scatter values in A). Significance was adjusted using the [13] method. **D-F**) show local Moran's $I$ values calculated based on the log-transformed counts of *KRT17* in a subsection of the CosMx human non-small cell lung cancer dataset [43]. A subset of the data is shown for illustration. The weight matrix was constructed using contiguous based neighbours in D), 10 nearest neighbours in E) and neighbours within a 1000 pixel distance ($\sim$ 180 µm), axes labels correspond to pixels in F). Axes values correspond to pixels with arbitrary origin.

## ***pasta*** highlights various spatial statistics analyses with inline code in R and Python

The collection of vignettes on our webpage (`https://robinsonlabuzh.github.io/pasta`) showcases the analysis of data from multiple spatial omics technologies, where concepts and assumptions are discussed in detail, with inline R and Python code. The design of the webpage follows the outline of Figure 1. First, we consider the technology and categorize into a HTS-based or imaging-based approach. Second, we distinguish between univariate (one type of mark) and multivariate (more than one type of mark) methods and outline different methods in specific vignettes with an in-depth discussion of the methods, e.g. providing mathematical definitions and formulae. In addition, we provide overview vignettes summarizing the two approaches.

To ensure usability, we showcase exploratory spatial statistics methods integrated into existing bioinformatics ecosystems. In Python, we use `anndata` objects from `scanpy`, allowing direct usage with `SpatialData` as well [86, 92, 52]. In R, we integrate with `SpatialExperiment` and `SpatialFeatureExperiment` objects from `Bioconductor` [72, 59, 33]. The packages `Voyager` and `spatialFDA` provide wrappers to compute both lattice and point pattern metrics on `Bioconductor` objects [59, 30]. Furthermore, our vignette offers examples how to convert between `anndata` objects in Python and `SpatialExperiment`, `Seurat objects` objects in R [72, 42]. Using these conversions, we are enabling analysis for users of both R and Python.

Overall, our resource Pattern Analysis for SpaTial omics datA (*pasta*) highlights the usefulness and transferability of existing exploratory spatial statistics approaches in the context of spatial tissue profiling.

## Spatial statistics is used in existing spatial omics data analysis tools

There are already several tools for omics analysis that employ spatial statistics. Among others, `Voyager` is a framework and a collection of use cases for lattice data written in both R and Python [59]; it provides data structures and functionalities to compute spatial statistics in molecular biology. `Voyager` comes with comprehensive vignettes for several spatial profiling technologies. The methodological focus of their vignettes is on lattice data analysis and the `spdep` geospatial package [15]. Since `Voyager` offers efficient implementations for lattice data analysis in R, we build on their framework in our vignettes for the lattice data component. Other methods written in R include: `SPIAT`, which contains various tools for spatial analysis, such as: neighbourhood analysis, local metrics and heterogeneity scores etc. [31]; `scFeatures`, a toolbox comprising lattice data analysis functionality such as Moran's $I$ and point pattern metrics as the $L$-function [21]; `spicyR`, which uses a compressed version of the global $L$-function for cross-sample comparison [20]; `lisaClust`, a spatial domain detection method via LISA (local indicators of spatial association) $L$ curves [65]; `mxfda`, that uses point pattern summary functions for survival analysis [93]; `spatialDM` uses global and local bivariate Moran's $I$ to score co-localization of ligand-receptor pairs [49]; and `MERINGUE`, which uses nearest-neighbour autocorrelation for spatially variable gene selection [54]. Methods in Python include `squidpy`, a package with various spatial statistics tools for both point pattern and lattice data [63]; `PySAL` a general Python library for spatial analysis [71]; `Bento` a toolkit for analysis of subcellular features using different spatial statistics [51]; `SpottedPy` a package designed to identify tumor hotspots using spatial statistics [91]; `MuSpAn` a toolbox for multiscale spatial analysis using spatial statistics [19] and `spatiomic` a Python library for the analysis of pixel based spatial data using spatial correlation measures [46].

For a detailed view on general methods development in the field of spatial transcriptomics, we refer readers to the online Supplementary table of the museum of spatial transcriptomics paper [58].

## Short guide to spatial analysis of spatial omics data

### Step 1 – The data representation

First, analysts should consider their technology and what options they have (i.e., imaging-based or HTS-based approaches). As seen in Figure 1, the data representation leads to either point pattern or lattice data analysis. If both streams are possible, the decision depends on the research question. In point pattern analysis, the process leading to the locations of individual points is the primary interest. It allows us to make statements such as whether a point process exhibits clustering or if two cell types tend to interact more than expected under CSR. In lattice data analysis, however, we are not directly interested in the arrangement of points, but rather locations are used to define the structure of the neighbourhood (i.e., via the weight matrix). We study the interaction of features, such as gene expression levels, given their spatial location.

### Step 2 – Choice of feature and nature of the spatial interaction

Second, the feature has to be considered. Most often this will be either gene or protein expression (continuous feature) or cell type or other labels (categorical feature). As seen in Figure 2, lattice data analysis offers many more methods for continuous features, whereas the focus of point pattern analysis is categorical features. Next, the scientific question guides the choice of the analysis. For example, if we are interested in the expression of one gene in space, this is a univariate continuous comparison. Another example would be the comparison of the co-localization of two cell types, which is a bivariate categorical comparison.

A fundamental choice in lattice data analysis is the construction of the weight matrix. As seen in Figure 5D-F, there are differences in the interpretation of results between different weight matrix choices. The appropriate weight matrix design depends on the research question. For instance, if the feature of interest is a ligand that requires direct cell contact, using contiguous neighbours is a reasonable choice. If the feature of interest is a molecule that diffuses to nearby cells (e.g., paracrine signalling), a nearest neighbour or distance-based approach should be considered for constructing the neighbourhood matrix.

Figure 1 helps with the navigation of both data modalities and analysis options.

### Step 3 – Observation scale

Third, the observation scale has to be chosen. Sometimes, a scale given by a technology such as a FOV might be appropriate. In other cases, the observation scale should be given by the biological question. Recall the example of Figure 4 where the scale effectively determines whether the point pattern of islet cells is considered to be homogeneous or clustered.

In lattice data analysis the scale is directly connected to the weight matrix. For example, if ligand-receptor co-localizations are of interest, the weight matrix can be specified in a way that captures spatial associations of gene expression between direct neighbours or at a specified scale. Generally, the concept of spatial autocorrelation is based on Tobler's famous first law of geography, which states that *"everything is related to everything else, but near things are more related than distant things"* [82]. This statement is not generally applicable in biology where we face highly compartmentalized anatomical structures. In analysis, it might therefore be important to account for this compartmentalization and define the neighbourhood accordingly or perform analysis that is aware of different compartments.

### Step 4 – Technical considerations

In point pattern analysis, the analyst needs to decide whether (and how) to control for inhomogeneous intensities, leading back to potential confounding of intensity and interaction [11, pp. 151 ff.]. Analysts can look at smoothed density estimates to inspect potential inhomogeneity prior to analyzing the pattern more formally. Furthermore, it depends again on the research question and biological domain knowledge whether the assumption of homogeneity is appropriate. While some tissues form regular structures that could be assumed to be homogeneous, other tissues appear to be more inhomogeneous. As shown in Figure 4, the homogeneous and inhomogeneous variants of the spatial summaries can lead to different interpretations.

Overall, we advise analysts to consider both scale and homogeneity carefully, keeping the biological context in mind and to document the decisions made and their rationale.

## Discussion

In this work, we presented multiple options for the exploratory analysis of spatial omics data. We introduced the foundations of both point pattern and lattice data analysis and showed the usability of both frameworks in accompanying vignettes.

Fundamentally, point pattern and lattice data analysis are two distinct paths to analyzing spatial omics data. Analysts should be aware of the assumptions and interpretations that can be drawn from each approach. In lattice data analysis, we regard the spatial coordinates as fixed and focus on the dependence between neighbours, i.e., weight matrices that reflect these neighbour interactions. In contrast, point pattern analysis concentrates on the underlying mechanisms that generate stochastic spatial patterns (i.e., clustered, regular or spaced arrangements). Moreover, the type of mark associated with spatial locations is another decisive factor in analysis. Lattice data analysis is most suited when the mark of interest is continuous, while point pattern analysis is the case for categorical marks. However, small nuances exist with lattice data analysis methods for categorical data and point pattern methods for continuous data. Therefore, the choice of the analysis framework depends not only on the data modalities, but also on the research question [5, 24, 11, 66, 25]. Meanwhile, there are dualities between the data modalities from spatially-resolved measurements (c.f., Figure 1). Using cell segmentation, imaging data can be approximated as points via their centroids. Point locations are stochastic and can therefore be regarded as a realization of a point process and can be studied using point pattern analysis. On the other hand, we can also view segmentation as an irregular lattice (and aggregate expression per cell). Similarly, HTS-based data with subcellular resolution (e.g., Visium HD [60]) can likewise be interpreted as a regular lattice of (sub)-cellular locations or irregular lattice from segmentation [68] or the reconstructed cells can then be approximated by their centroids as points and further analyzed using point pattern analysis.

We have focused here on spatial statistics and highlighted two main streams of analysis. Another popular way to represent spatial arrangements is as spatial graphs [2, 90]. The methods described in the section on lattice data are closely connected to spatial graphs, especially in the construction of a weight matrix. Point pattern analysis concepts are also related to spatial graphs via edge rules. For example, Ripley's $K$ can be interpreted as a graph problem with a distance threshold edge rule [25, p. 136].

## Finding the correct scale of analysis

One common challenge in all spatial analyses is the question of the correct scale. In Figure 4, we showed that the scale influences the interpretation of the spatial analysis: what is found to be spatially heterogeneous at a one scale can be locally homogeneous at another [84, 25]. Thus, the scale should be defined in accordance with the scientific question.

Moreover, the definition of cells and spatial structures or domains is often not straightforward and in turn leads to challenges that can be summarized as the modifiable areal unit problem (MAUP) [61, 38], which states that the definition of a region will affect downstream analyses and conclusions. For more discussion about the MAUP in spatial transcriptomics, we refer the reader to Zormpas et al.

## Limitations of exploratory spatial statistics

There are various other limitations when studying spatially-resolved omics data. On the technical side, imperfect sections can lead to artifacts or non-overlapping FOVs, which can introduce missing data. Furthermore, cells are never a perfectly monolayer within a histological section, thus cells overlapping in the $z$-axis are to be expected. In terms of analysis, two-dimensional views neglect processes that occur in three dimensions. For example, in lattice data we might neglect an important contiguous neighbour in an adjacent slice. The same applies to point pattern analysis where we consider only processes in two dimensions even though the underlying biological process likely happens in three dimensions. Both lattice data and point pattern analysis extend conceptually to 3D, however with increased computational complexity. Moreover, in point pattern analysis, we often implicitly assume that patterns are invariant to rotations (see Box Point Processes). This assumption is not anatomically accurate for organs and tissues with layered structures, such as the brain. Moreover, there are other limitations arising from the concept of spatial metrics. Lattice data analysis methods were originally designed for rather low dimensional datasets, while spatial omics often deals with thousands of features (e.g., untargeted transcriptome). In addition, the interpretation of a spatial correlation metric depends on the measure of similarity that is compared in a local neighbourhood. In a multivariate setting, a straightforward interpretation of this similarity measure is often not possible. Similarly, point pattern methods usually only allow for pairwise comparison of features, which can lead to a high number of cross comparisons. Moreover, with newer technologies, the size of the biological sample and resolution of measurements are constantly expanding, which can be computationally expensive. Since point pattern based methods only store coordinates, they usually scale better in terms of memory requirements.

In our accompanying vignette, we discuss concepts and assumptions in more detail using biological examples. For point pattern analysis, we mainly use R due to the lack of comprehensive Python libraries available for point pattern analysis. Since the R package `Voyager` is an extensive framework and resource for lattice data, we build on it for the lattice data analysis part of our work. In addition, we provide vignettes for point pattern analysis to complement existing discussions [95, 59].

### Future opportunities

The spatial metrics as described in point pattern and lattice data analysis offer means to quantify spatial patterns in molecular biology. There are still open research questions including the correct application (e.g., scale and homogeneity) of methods to a given biological context, the high dimensional nature of spatial omics data and the effect of preprocessing steps that come upstream of the application of spatial statistics. Furthermore, comparisons of spatial metrics across multiple samples and conditions and parametric spatial models offer interesting options for future work.

## Methods

All analysis were conducted using R version 4.5.0 and Bioconductor version 3.21 [33]. Results were visualized using `ggplot2` [89].
The Janesick et al. Xenium breast cancer dataset was downloaded using the package `STexampleData` [88], cell type labels and the H&E image (licensed under CC-BY 4.0) were downloaded from the 10x website. Gene expression normalization was performed using the function `logNormCounts` of the `scuttle` package [53]. Spatial-aware normalization was conducted as described in the package `SpaNorm` [76]. The 10x Visium mouse brain dataset [1] was downloaded using the package `STexampleData`. Quality control metrics were calculated using the package `scuttle` and gene expression normalization was performed using the function `logNormCounts` of the `scuttle` package. Preprocessing was conducted as described in the OSTA book `https://lmweber.org/OSTA/`. The He *et al.* CosMx non-small lung cancer dataset [43] was downloaded using the `SFEData` package [59]. Preprocessing was conducted as described in the `Voyager` online vignette [59]. The Damond *et al.* IMC dataset [27] was downloaded using the package `imcdatasets` [26], cell type labels were used as provided by the authors.
The mouse brain datasets shown in Figure 1 (Technology) were taken from Shi *et al.* (STARMAP Plus) [77] [77] and 10x (Visium mouse) [1]. Clustering was performed using `Banksy` [78].

Lattice data analysis was performed using the package `Voyager` [59] and `spdep` [16], point pattern analysis was conducted using the packages `spatialFDA` [30] and `spatstat` [12]. Biologically relevant observation windows were determined using the package `sosta` [40].

## Code and data availability

Code and instructions to retrieve data to reproduce figures and other results can be found under `https://github.com/robinsonlabuzh/pasta-manuscript`, code licensed under the GPLv3 terms. The vignette and source code can be accessed via `https://robinsonlabuzh.github.io/pasta` licensed under CC-BY-4.0 terms for the text and GPLv3 license for any code.
Versioned source code and rendered vignettes available under `https://doi.org/10.5281/zenodo.15488040`.

## Acknowledgments

We thank all members of the Robinsonlab for constructive feedback on the manuscript and vignette. In particular, we thank Peiying Cai, Alice Driessen, Reto Gerber, Pierre-Luc Germain, Maruša Kodermann, Vladyslav Korobeynyk, Siyuan Luo, Giulia Moro, Emanuel Sonder, Jiayi Wang and David Wissel for their careful reading and input to the vignette.The Python examples in the vignette were developed as part of the SpaceHack 2024 event. We thank the SpaceHack organisers and participants. Furthermore, we would like to thank Daria Lazic for helpful comments on the manuscript.

## Author contributions

**ME:** Conceptualisation, Methodology, Software (Vignette), Analysis, Visualisation, Writing - Original Draft; **SG:** Conceptualisation, Methodology, Software (Vignette), Analysis, Visualisation, Writing - Original Draft; **HLC:** Conceptualisation, Vignette, Writing - Review & Editing; **IM:** Methodology, Software (Vignette), Writing - Review & Editing; **MK**, Software (Vignette), Writing - Review & Editing; **RF:** Methodology, Writing - Review & Editing; **MDR:** Conceptualisation, Methodology, Writing - Original Draft, Supervision, Funding acquisition.

## Funding

This work was supported by Swiss National Science Foundation (SNSF) project grant 310030_204869 to MDR. MDR acknowledges support from the University Research Priority Program Evolution in Action at the University of Zurich. HLC acknowledges support by SNSF grant number 222136.

## Conflict of interest

We declare no conflict of interest.

## Supplementary data

Table S1: Non-exhaustive table of technologies and the data modality they can represent. For example, MERFISH can be represented as an irregular lattice using the outline of the cell segmentations. The cell centroids of the cell segmentations as well as the transcript locations can be represented as a point pattern.

| **Technology** | **Main feature of interest** | **Lattice data** | **Point pattern** |
|---|---|---|---|
| Visium [79] | transcriptomics | spots (regular) | not applicable |
| Slide-seq V1& V2 [75, 80] | transcriptomics | beads (regular);<br>segmented cells (irregular) | segmented cells |
| Visium HD [60] | transcriptomics | spots (regular);<br>segmented cells (irregular) | segmented cells |
| Stereo-seq [23] | transcriptomics | spots (regular);<br>segmented cells (irregular) | segmented cells |
| Xenium [44] | transcriptomics | segmented cells (irregular) | transcript locations;<br>segmented cells |
| CosMx [23] | transcriptomics | segmented cells (irregular) | transcript locations;<br>segmented cells |
| MERFISH [23] | transcriptomics | segmented cells (irregular) | transcript locations;<br>segmented cells |
| IMC [36] | proteomics | pixels (regular);<br>segmented cells (irregular) | segmented cells |
| MIBI-TOF [45] | proteomics | pixels (regular);<br>segmented cells (irregular) | segmented cells |
| CODEX [17] | proteomics | segmented cells (irregular) | transcript locations;<br>segmented cells |
| 4i [41] | proteomics | segmented cells (irregular) | transcript locations;<br>segmented cells |

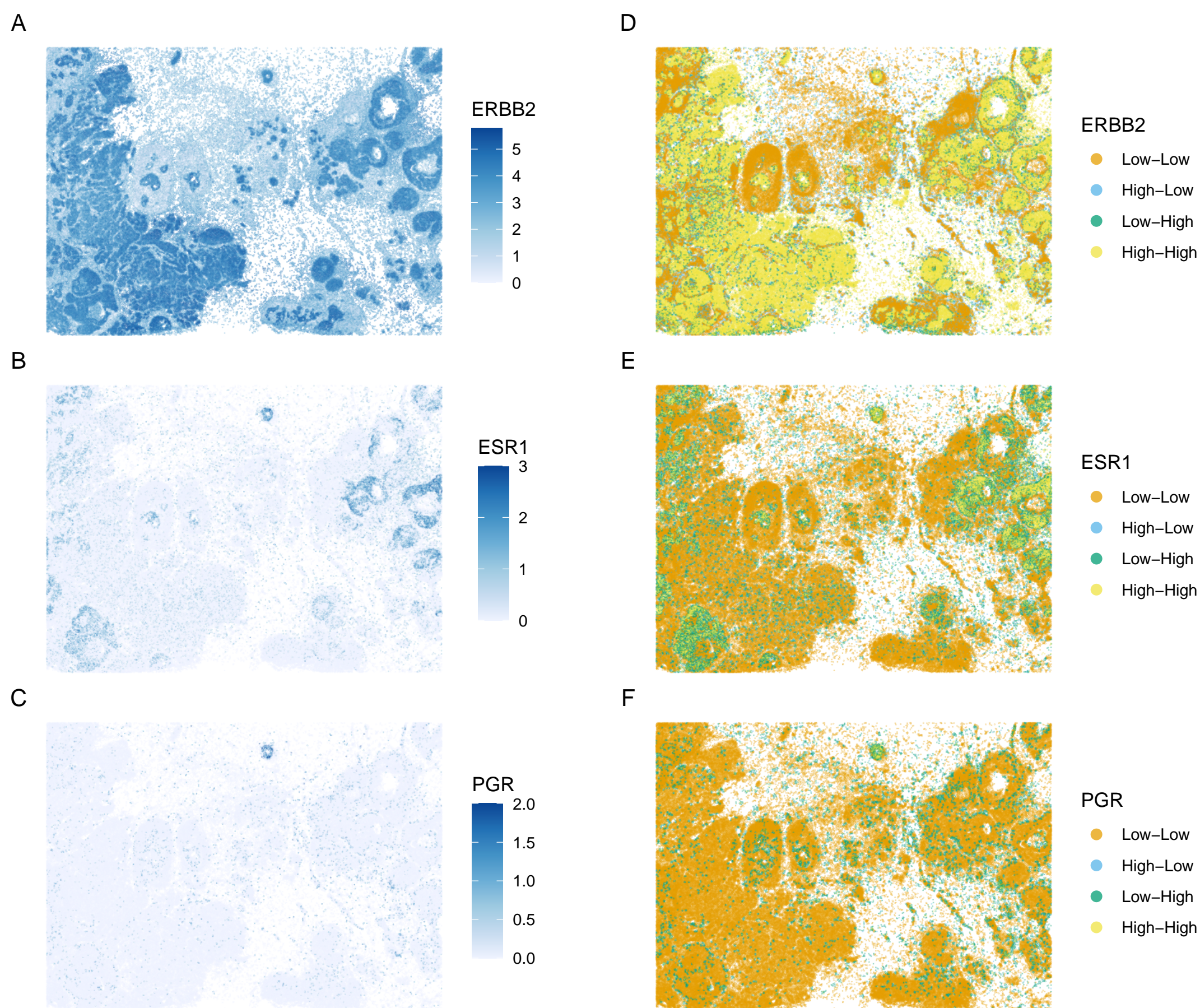

Figure S1: Spatial correlation (lattice data) analysis of three receptor genes (*ERBB2*, *ESR1* and *PGR*) in a Xenium human breast cancer data set [44]. **A-C**) The gene expression of three receptor genes *ERBB2*, *ESR1* and *PGR*. The neighbourhood of a location is defined as the 6 nearest neighbours. **D-F**) Moran's scatter plot to assess the type of spatial autocorrelation of the three genes.

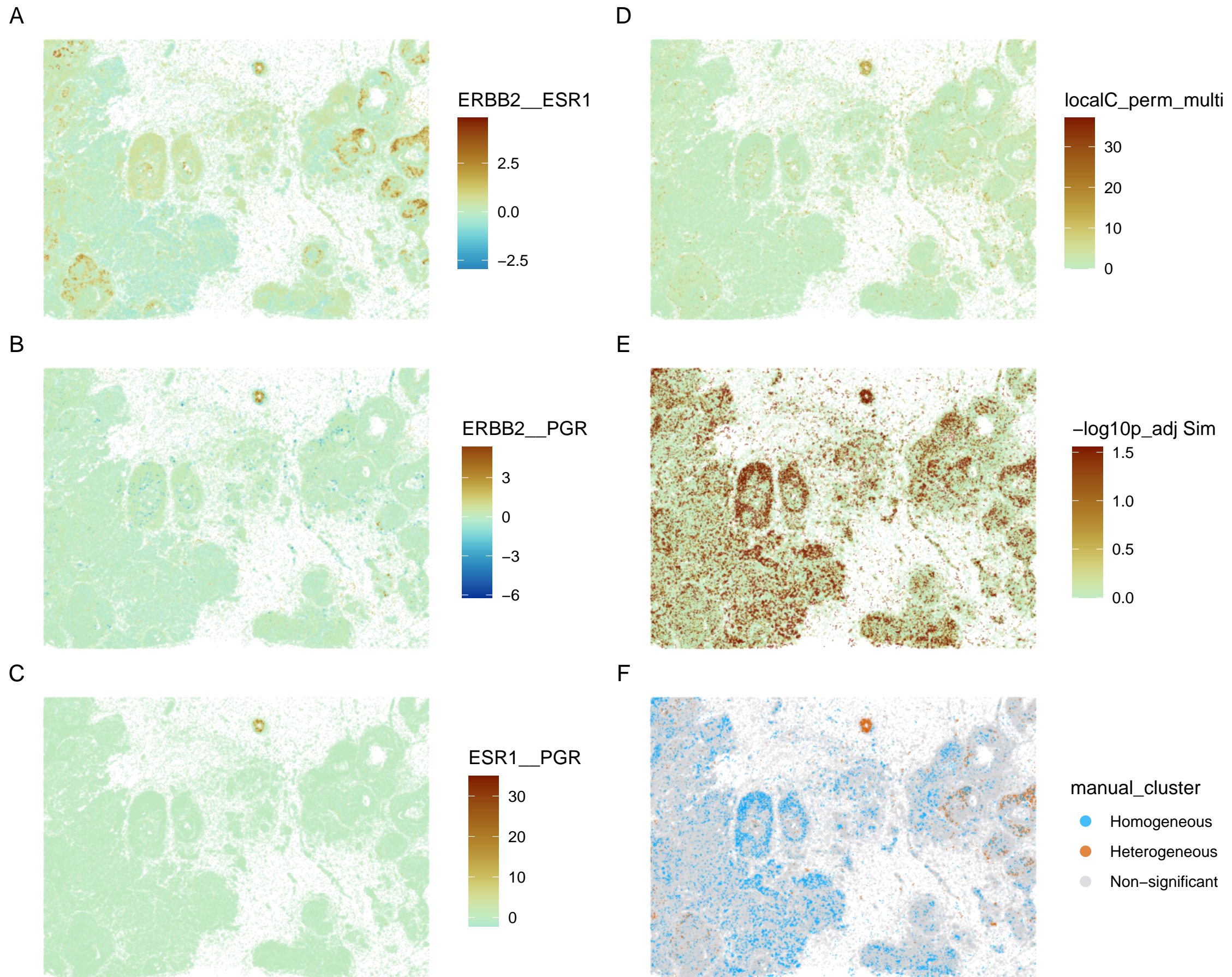

Figure S2: **A-C**) Bivariate Lee's $L$ assessing spatial correlation for the pairwise combinations of the three genes. **D**) Multivariate Geary's $c$ assessing spatial correlation for the three genes. Positive values indicate local heterogeneity. **E**) The corresponding adjusted p-values to the multivariate Geary's $c$ on a log scale obtained from permutations. **F**) Multivariate Geary's $c$ values were defined as homogeneous if $c < 1$ and heterogeneous if $c \geq 1$. Significance was determined by adjusted $p < 0.05$

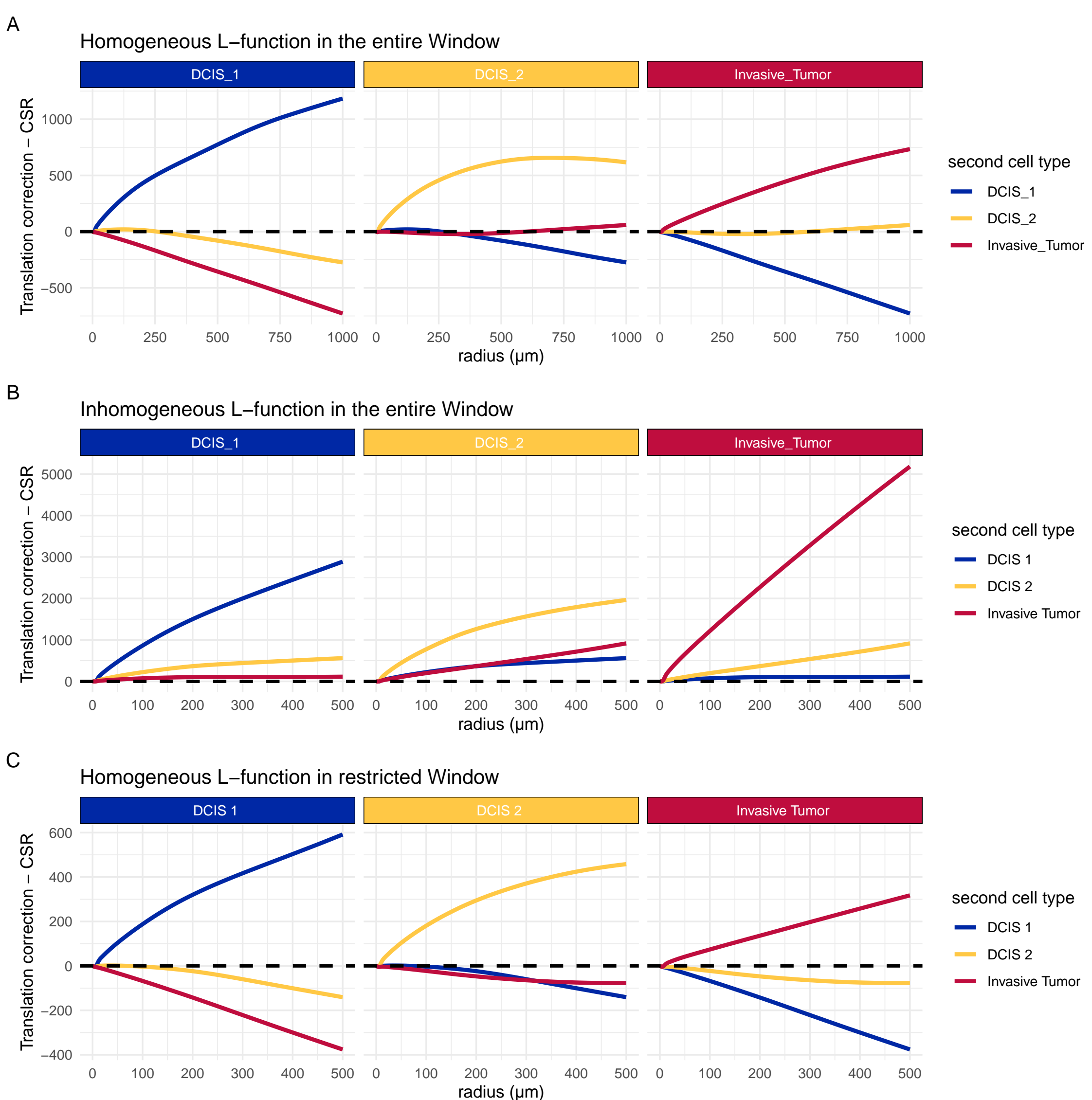

Figure S3: Three point pattern analyses of a Xenium human breast cancer data set [44]. Analysed are ductual carcinoma in situ 1 and 2, as well as invasive tumour cells. **A**) Homogeneous Besag's $L$ function for the pairwise cell type combinations in the entire window, not correcting for any inhomogeneity in the distribution of cells in the tissue. **B**) Inhomogeneous Besag's $L$ function for the pairwise cell type combinations in the entire window scaled by the average intensity of the unmarked point pattern per unit square. This corrects for the underlying inhomogeneous distribution of cells in the tissue. **C**) Homogeneous Besag's $L$ function for the pairwise cell type combinations in a restricted window. The restricted window was determined with an intensity threshold on the unmarked point pattern.

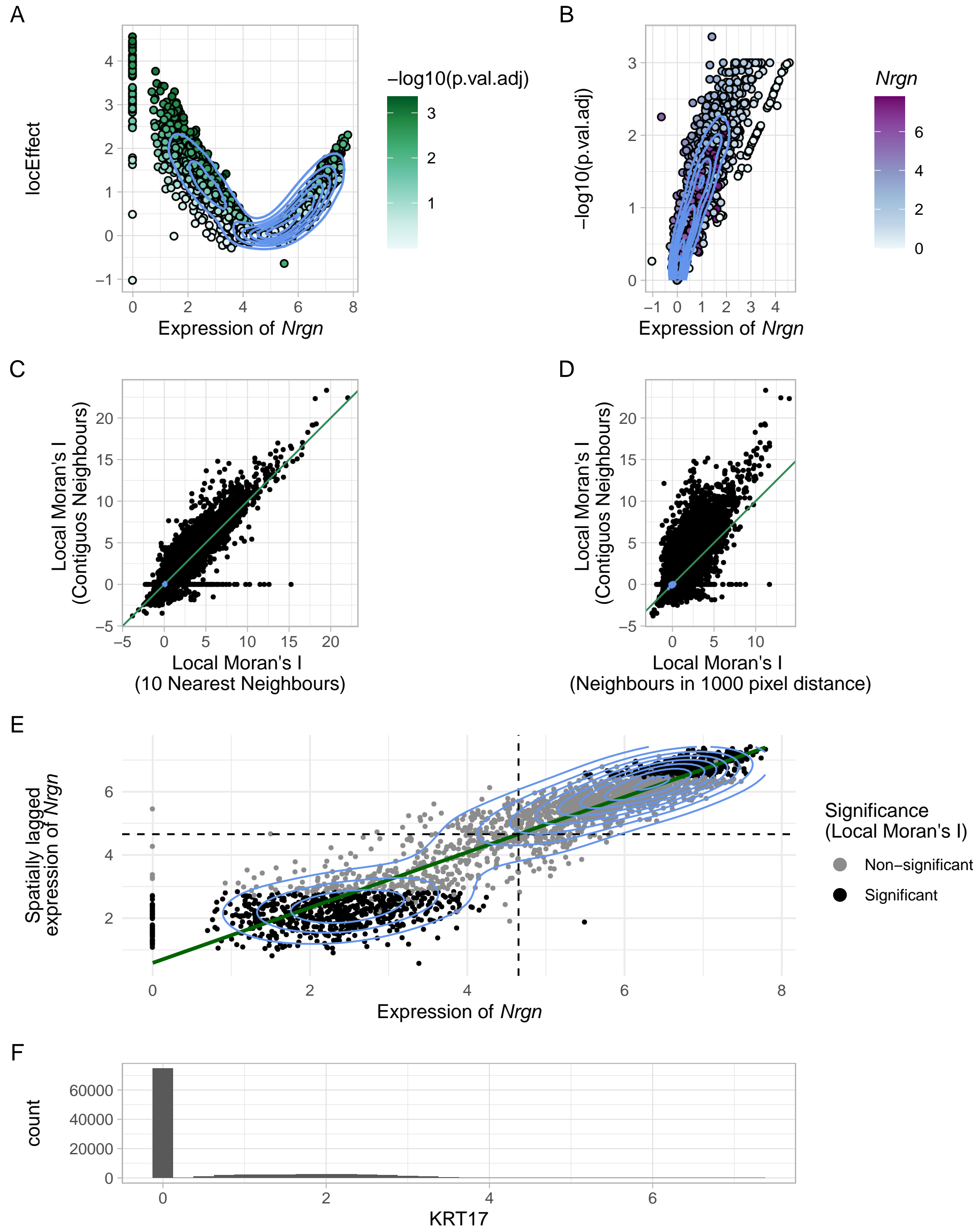

*(Figure legend on next page.)*

Figure S4: In **A-B**) each point represents a spot in the Visium mouse brain dataset [1]. Dependence between log-transformed counts of gene *Nrgn* and the corresponding local Moran's $I$ values in A); the corresponding local Moran's $I$ and adjusted p-values in B). The colours indicate the corresponding adjusted p-value in A) and log-transformed counts in B). **C-D**) show the relationship of local Moran's $I$ values and the log-transformed counts of gene *KRT7* in the CosMx human non small cell lung cancer dataset [43]. Local Moran's $I$ values calculated based on contiguous based neighbours vs. distance based neighbours in C); and contiguous based neighbours vs. neighbours in 1000 pixel distance in D). Red indicates cells with no contiguous neighbours. Green line indicates $x = y$. Blue lines indicate local densities. Note the high density close to the origin in C) and D) resulting from sparse expression of the gene *KRT7*, c.f., histogram in **F**). **E)** Moran's scatter plot displaying the expression of gene *Nrgn* in each spot vs. the average expression of its neighbours (as defined in the weight matrix). Significance values correspond to local Moran's $I$. The dotted lines correspond to the mean expression of *Nrgn* and are used to define clusters; c.f., Figure 5 C).

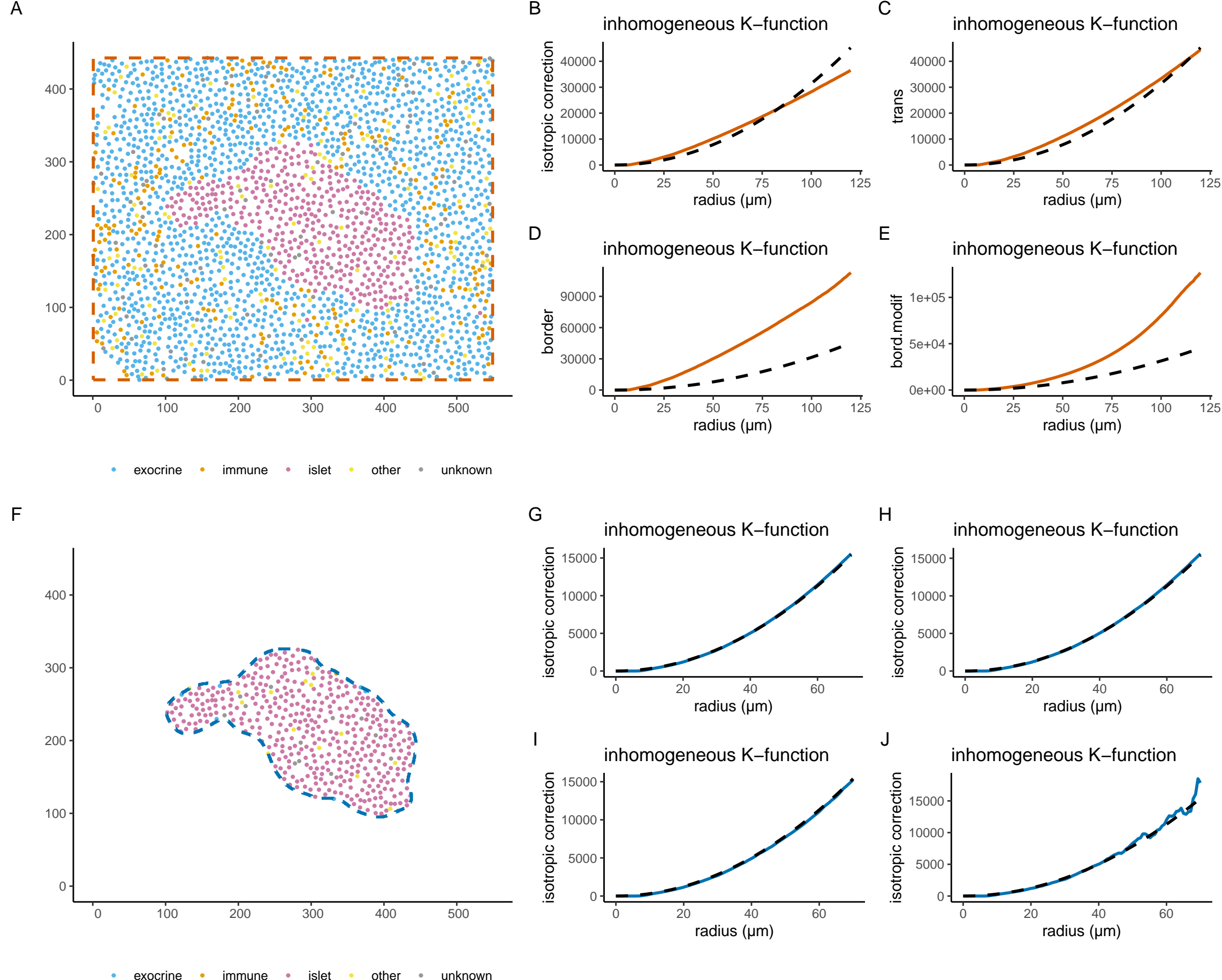

Figure S5: **A**) Single image of an IMC dataset showing islets in the human pancreas [27] The analysis window (dashed line) is set to correspond to the entire FOV. Global analysis using inhomogeneous $K$-function with **B**) isotropic, **C**) translational, **D**) border and **E**) modified border correction. **F**) Local analysis on subset of islet cells and window (dashed line) around tissue structure using inhomogeneous $K$-function with **G**) isotropic, **H**) translational, **I**) border and **J**) modified border correction.